\documentclass[pre,twocolumn,amsmath,amssymb,nofootinbib,floatfix,superscriptaddress]{revtex4}

\usepackage{graphicx,bm,xcolor, bbold}
\usepackage{enumerate}
\usepackage{graphicx,bm}
\usepackage[normalem]{ulem}

\makeatletter
\def\graphicscale{\twocolumn@sw{0.3}{0.4}}
\def\graphicthreescale{\twocolumn@sw{0.3}{0.4}}

\begin{document}

\title{Critical relaxational dynamics at the continuous
  transitions of \\ three-dimensional  
  spin models with ${\mathbb Z}_2$ gauge symmetry}

\author{Claudio Bonati} \affiliation{Dipartimento di Fisica
  dell'Universit\`a di Pisa and INFN Sezione di Pisa, Largo Pontecorvo 3,
  I-56127 Pisa,
  Italy}

\author{Andrea Pelissetto}
\affiliation{Dipartimento di Fisica dell'Universit\`a di Roma Sapienza
        and INFN Sezione di Roma I, I-00185 Roma, Italy}

\author{Ettore Vicari} 
\affiliation{Dipartimento di Fisica dell'Universit\`a di Pisa,
        Largo Pontecorvo 3, I-56127 Pisa, Italy}

\date{\today}

\begin{abstract}
  We characterize the dynamic universality classes of a relaxational
  dynamics under equilibrium conditions at the continuous transitions
  of three-dimensional (3D) spin systems with a ${\mathbb Z}_2$-gauge
  symmetry. In particular, we consider the pure lattice ${\mathbb
    Z}_2$-gauge model and the lattice ${\mathbb Z}_2$-gauge XY model,
  which present various types of transitions: topological transitions
  without a local order parameter and transitions characterized by
  both gauge-invariant and non-gauge-invariant XY order parameters.
  We consider a standard relaxational (locally reversible) Metropolis
  dynamics and determine the dynamic critical exponent $z$ that
  characterizes the critical slowing down of the dynamics as the
  continuous transition is approached.  At the topological ${\mathbb
    Z}_2$-gauge transitions we find $z=2.55(6)$. Therefore, the
  dynamics is significantly slower than in Ising systems---$z\approx
  2.02$ for the 3D Ising universality class---although 3D ${\mathbb
    Z}_2$-gauge systems and Ising systems have the same static
  critical behavior because of duality.  As for the nontopological
  transitions in the 3D ${\mathbb Z}_2$-gauge XY model, we find that
  their critical dynamics belong to the same dynamic universality
  class as the relaxational dynamics in ungauged XY systems,
  independently of the gauge-invariant or nongauge-invariant nature of
  the order parameter at the transition.
\end{abstract}

\maketitle

\section{Introduction}
\label{intro}

Gauge symmetries are key features of theories describing the
fundamental interactions~\cite{Weinberg-book,ZJ-book} and some
collective phenomena in condensed-matter
physics~\cite{Anderson-book,Wen-book,Fradkin-book,Sachdev-book2}.
Several collective phenomena can be modelled by effective lattice
gauge theories in which scalar variables are coupled with gauge fields
associated with discrete (such as ${\mathbb Z}_2$) or continuous [such
  as U(1) or SU($N$)] gauge groups.  The nature of their different
phases and phase transitions crucially depends on the interplay
between global and gauge symmetries, and, in particular, on the role
played by the gauge modes.  Many lattice Abelian and non-Abelian gauge
models have been considered, with the purpose of identifying the
possible universality classes of their continuous transitions, see,
e.g., Refs.~\cite{BPV-25,Sachdev-book2,Sachdev-19} and references
therein.  In some cases these models show phase transitions that
cannot be described by Landau-Ginzburg-Wilson (LGW) $\Phi^4$
theories~\cite{Wilson-83,ZJ-book,PV-02} with a local gauge-invariant 
order-parameter field, see, e.g., Refs.~\cite{SBSVF-04, Sachdev-19,
  Senthil-23, BPV-25}.  These non-LGW transitions are driven by
extended charged excitations with no local order parameter or by a
nontrivial interplay between long-range scalar fluctuations and
nonlocal topological gauge modes.

The classical (thermal) continuous transitions of lattice gauge
systems can be generally classified into four broad classes, each one
admitting a different type of effective description~\cite{BPV-25}: (i)
LGW transitions with a gauge-invariant order parameter and noncritical
gauge modes, which can be described by a standard LGW $\Phi^4$ theory
for a gauge-invariant order-parameter field; (ii) LGW$^{\times}$
transitions, in which gauge modes are also not critical, but the
order-parameter field of the corresponding effective LGW $\Phi^4$
theory is not gauge invariant; (iii) Gauge-field theory (GFT)
transitions, where also gauge modes are critical, thus requiring an
effective field-theoretical description that includes the gauge
fields; (iv) Topological transitions driven by topological gauge
modes, without any local gauge-invariant scalar order parameter.
Clearly, critical phenomena in the presence of gauge symmetries show a
more complex phenomenology than standard spin systems, reflecting the
fact that, in addition to the usual disordered (high-temperature) and
ordered (low-temperature) phases, uniquely characterized by the
behavior of the matter fields, these systems also admit Higgs and
topological phases~\cite{Wen-book, Fradkin-book, Sachdev-book2,
  Sachdev-19, BPV-25}.  Analytical and numerical results for
three-dimensional (3D) lattice gauge models support the above
classification, see, e.g.,
Refs.~\cite{BPV-25,Wegner-71,HLM-74,BDI-74,OS-78,FS-79,Kogut-79,Hikami-80,
  DH-81,CC-82,FM-83, KK-85,KK-86,BN-87,MS-90,
  LRT-93,KKS-94,BFLLW-96,HT-96,FH-96,IKK-96,KKLP-98,OT-98,
  SF-00,SSS-02,SSSNH-02,KNS-02,MHS-02,SM-02,NRR-03,SSNHS-03,MZ-03,KS-08,
  CAP-08,TKPS-10,BMK-13, KNNSWS-15,NCSOS-15,WNMXS-17,
  PTV-18,IZMHS-19,PV-19-AH3d,SSST-19,BPV-19,SPSS-20,BPV-20-hcAH,BPV-20-on,
  BPV-21-ncAH,SSN-21,BFPV-21-su-ad,BFPV-21-su,
  BPV-22,BPV-22-z2h,BPV-23-gf,BP-23,BF-23,ZZV-23,
  BPV-23-chgf,BPV-23-mppo,BPV-24-z2gaugeN,
  BPV-24-onstar,XPK-24,BPV-24-decQ2,SSN-24,BPV-24-ncAH,BPSV-24}.

Most studies have so far focused on the equilibrium static properties,
determining the phase diagrams and characterizing the critical
properties of the continuous transitions. Instead, little is known of
the critical dynamics in gauge systems. Indeed, while the critical
features of the dynamics have been extensively investigated in systems
without gauge symmetries (see, e.g., Ref.~\cite{Ma-book,HH-77,FM-06}
and references therein), only few studies have considered the dynamics
in statistical systems with gauge symmetries. As far as we know, there
are only some numerical studies of the relaxational dynamics in the 3D
${\mathbb Z}_2$-gauge model without matter
fields~\cite{BKKLS-90,XCMCS-18}, some studies of the dynamic
properties of superconductors (see, e.g., Refs.~\cite{DFM-07, LVF-04,
  SBZ-02, AG-01, JKM-00, LWWGY-98, WJ-97, LMG-91}) and of hadronic
matter close to continuous phase transitions~\cite{RW-93}. We also
mention that the purely relaxational dynamics of gauge field theories
has also been addressed in the context of the stochastic quantization
of gauge theories, see, e.g.,
Refs.~\cite{PW-81,Zwanziger-81,FI-82,NNOO-83,ZJ-86,ZZ-88}.  We believe
that further studies are needed to obtain a deeper understanding of
the critical dynamics in the presence of gauge symmetries. In
particular, they are needed to understand how the different peculiar
features of the static critical behavior affect the dynamic behavior.

In this paper we begin addressing these issues by studying the purely
relaxational dynamics in the presence of a local ${\mathbb Z}_2$-gauge
symmetry. For this purpose, we consider the topological
finite-temperature transition in the 3D lattice ${\mathbb Z}_2$-gauge
model~\cite{Wegner-71} and the continuous transitions in the 3D
${\mathbb Z}_2$-gauge $N$-vector model~\cite{BPV-24-z2gaugeN} (we
focus on the XY case $N=2$), in which an $N$-vector field is minimally
coupled to ${\mathbb Z}_2$ link variables.

The 3D lattice ${\mathbb Z}_2$-gauge theory is a paradigmatic model
undergoing a finite-temperature topological
transition~\cite{Wegner-71,Sachdev-19} without any local order
parameter, separating the high-temperature deconfined phase from the
low-temperature confined phase.  By duality, this model can be related
to the standard Ising model in the absence of an external magnetic
field, implying that energy observables have the same critical
behavior in the two models.

\begin{figure}[tbp]
\includegraphics[width=0.9\columnwidth, clip]{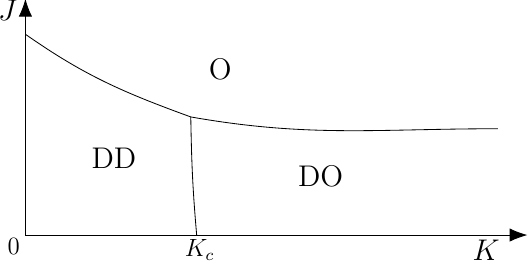}
\caption{Sketch of the phase diagram of the 3D ${\mathbb Z}_2$-gauge
  $N$-vector models for $N\ge 2$, in the space of the Hamiltonian
  parameters $K$ and $J$, cf. Eq.~(\ref{ham}). For small $J$ there are
  two spin-disordered phases: a small-$K$ phase, in which both the
  spins and the gauge variables are disordered (DD phase), and a
  large-$K$ phase, in which the ${\mathbb Z}_2$-gauge variables order
  (DO phase). In the large-$J$ phase both spins and gauge variables
  are ordered (O phase). The critical behavior along the three
  transition lines is discussed in
  Refs.~\cite{BPV-24-z2gaugeN,BPV-24-onstar,BPV-25}.  For $N=2$ (gauge
  XY model), DD-O transitions are LGW transitions, the order parameter
  being a gauge-invariant operator that is bilinear in the spin
  variables; DO-O transitions are LGW$^\times$ transitions, the order
  parameter being the gauge-dependent spin
  variable~\cite{BPV-24-onstar}; DD-DO transitions are topological in
  the ${\mathbb Z}_2$-gauge universality class.  The three transition
  lines meet at $(K_\star\approx 0.75,J_\star\approx 0.23)$.  }
\label{phadiaN}
\end{figure}

The phase diagram of the 3D ${\mathbb Z}_2$-gauge XY model shows
different phases characterized by the spontaneous breaking of the
global SO(2) symmetry and by the different topological properties of
the ${\mathbb Z}_2$-gauge correlations, see, e.g.,
Refs.~\cite{FS-79,Sachdev-19,BPV-24-z2gaugeN}.  As sketched in
Fig.~\ref{phadiaN}, two spin-disordered phases are present for small
$J$: a small-$K$ phase, in which both spin and ${\mathbb Z}_2$-gauge
variables are disordered (DD phase), and a large-$K$ phase in which the
${\mathbb Z}_2$-gauge variables order (DO phase).  For large $J$ there is a
single phase in which both spins and gauge variables order (O phase).  The
transitions along the three lines separating the DD, DO,
and O phases have different features and belong to different classes,
according to the classification reported above~\cite{BPV-24-z2gaugeN}:
LGW, LGW$^\times$, and topological transitions occur along the DD-O,
DO-O, and DD-DO transition lines~\cite{BPV-25}, respectively. The
topological transitions along the DD-DO line belong to the 3D
${\mathbb Z}_2$-gauge universality class, as the topological
transition in the pure lattice ${\mathbb Z}_2$-gauge model.

In the present work we numerically study the purely relaxational
dynamics (corresponding to the model A in the classification reported
in Ref.~\cite{HH-77}) in these lattice ${\mathbb Z}_2$-gauge models,
to understand how the different nature of the transitions affects the
dynamics.  Specifically, we consider a standard, locally reversible
Metropolis dynamics, as commonly used in Monte Carlo (MC)
simulations~\cite{Binder-76}, focusing on the behavior under
equilibrium conditions.  We determine autocorrelation times $\tau$ of
gauge-invariant observables associated with the critical modes. In the
infinite-volume limit, they diverge as $\tau\sim\xi^z$
approaching the critical point, where $\xi$ is a diverging correlation
length scale and $z$ is a universal dynamic exponent, while, in a
finite volume and at the critical point, they diverge as $\tau\sim
L^z$, where $L$ is the linear size of the system.

As we shall see, the relaxational dynamics at the topological
transitions of the 3D pure ${\mathbb Z}_2$-gauge model and of the 3D
${\mathbb Z}_2$-gauge XY model turns out to be significantly slower
than that of the standard Ising model, although the static critical
behavior is the same (at least in the thermal sector).  Indeed, we
estimate $z = 2.55(6)$ for the topological transitions, that is
significantly larger than the value~\cite{Hasenbusch-20} $z=
2.0245(15)$ for Ising systems.  It is important to note that the
duality mapping between the ${\mathbb Z}_2$-gauge and the standard
Ising model that guarantees the equivalence of the static critical
behavior is nonlocal. Therefore, a local dynamics in the Ising model
would correspond to a nonlocal dynamics in the ${\mathbb Z}_2$-gauge
model. It is thus not surprising that a local dynamics gives rise to
two different dynamic universality classes for the ${\mathbb Z}_2$-gauge 
and the standard Ising model.

Concerning the continuous LGW and LGW$^\times$ transitions in the 3D
${\mathbb Z}_2$-gauge XY model, our numerical results show that the
universal features of the relaxational dynamics for these transitions
are the same as in XY systems without gauge invariance, with the same
dynamic exponent $z\approx 2.022$.  At LGW transitions, it implies
that the Metropolis dynamics defined on the lattice gauge-dependent
variables is effectively equivalent to the model-A dynamics
\cite{HH-77} in an effective LGW $\Phi^4$ model in which the
fundamental field is a composite gauge-invariant operator and no gauge
fields are present. At LGW$^\times$ transitions, it implies that both the static
and the dynamic behavior are insensitive to the presence of gauge
fields.

The paper is organized as follows.  In Sec.~\ref{model} we define the lattice
${\mathbb Z}_2$-gauge model and the ${\mathbb Z}_2$-gauge $N$-vector model, and
summarize the main features of their phase diagrams and transitions.
In Sec.~\ref{anreldyn} we discuss the definition of the
autocorrelation times we use to determine the dynamic critical exponent $z$,
and the definitions of the relevant observables. In Sec.~\ref{reldynz2gau} we
present a numerical study of the critical relaxational dynamics of the
${\mathbb Z}_2$-gauge model, considering a standard Metropolis dynamics for the
gauge variables. In Sec.~\ref{reldynz2nvec} we report our numerical analysis of
the purely relaxational dynamics at the continuous transitions of the 3D
${\mathbb Z}_2$-gauge XY model, along the DD-O, DO-O, and DD-DO transitions
lines, see Fig.~\ref{phadiaN}, which allows us to determine the universal
dynamic exponent $z$ in all cases. In Sec.~\ref{conclu} we summarize the
results and draw our conclusions. Finally, some numerical results are reported
in App.~\ref{data}.

\section{Lattice ${\mathbb Z}_2$-gauge models}
\label{model}

\subsection{Pure ${\mathbb Z}_2$-gauge theory}
\label{z2gaugem}

The 3D lattice ${\mathbb Z}_2$-gauge model is a paradigmatic model
undergoing a finite-temperature topological
transition~\cite{Wegner-71,Sachdev-19}.  On a cubic lattice, its
Hamiltonian reads
\begin{equation}
H_G(K) = - K \sum_{{\bm
      x},\mu>\nu}
   \sigma_{{\bm
      x},\mu} \,\sigma_{{\bm x}+\hat{\mu},\nu} \,\sigma_{{\bm
       x}+\hat{\nu},\mu} \,\sigma_{{\bm x},\nu},
   \label{Hgz2}
\end{equation}
where $\sigma_{{\bm x},\mu}=\pm 1$ is a link variable associated with
the bond starting from site ${\bm x}$ in the positive $\mu$ direction,
$\mu=1,2,3$.  The Hamiltonian parameter $K$ plays the role of inverse
gauge coupling, therefore the $K\to\infty$ limit corresponds to the
small gauge-coupling limit. In the following we set the temperature
$T=1$, thus the partition function reads
$Z(K)=\sum_{\{\sigma\}} e^{-H_G(K)}$.

The lattice ${\mathbb Z}_2$-gauge model undergoes a continuous
transition separating a high-$K$ deconfined phase from a low-$K$
confined phase. It is related to the standard 3D Ising model by a
duality mapping that relates the partition functions of the two
models~\cite{Savit-80,Wegner-71}. Duality implies that thermal
observables have the same critical behavior in the two models and, in
particular, that the correlation-length exponent $\nu$ and the
correction-to-scaling exponent $\omega$ are the same. In the following
we use the Ising estimates~\cite{KPSV-16} $\nu=0.629971(4)$ and
$\omega= 0.8297(2)$, see also
Refs.~\cite{Hasenbusch-21,FXL-18,KP-17,Hasenbusch-10,CPRV-02,GZ-98}.  Morever,
using the available estimates of the critical point for the standard
Ising model, see, e.g., Ref.~\cite{FXL-18}, duality allows us to
obtain an accurate estimate of the critical point of the model
(\ref{Hgz2}),
\begin{equation}
  K_c= 0.761413292(11).
  \label{Kcest}
  \end{equation}
As we shall
see below, the equivalence of the asymptotic critical behavior of
${\mathbb Z}_2$-gauge and Ising models does not extend to the critical
dynamics, essentially because the highly nonlocal duality mapping does
not extend to the relaxational dynamics.

We finally remark that the ${\mathbb Z}_2$-gauge transition is
topological, as it is not driven by a local order parameter. The
nonlocal Wilson loop $W_C$, defined as the product of the link
variables along a closed contour $C$ within a plane, provides a
nonlocal order parameter for the transition~\cite{Wegner-71}. Indeed,
its size dependence for large contours changes at the transition: It
obeys the area law $W_C\sim \exp(- c_a A_C)$, where $A_C$ is the area
enclosed by the contour $C$ and $c_a>0$ is a constant, for small
values of $K$ and the perimeter law $W_C\sim \exp(- c_p P_C)$, where
$P_C$ is the perimeter of the contour $C$ and $c_p>0$ is a constant,
for large values of $K$.

\subsection{The lattice ${\mathbb Z}_2$-gauge $N$-vector model}
\label{z2gaugeNvector}

The lattice ${\mathbb Z}_2$-gauge $N$-vector model is a paradigmatic
model describing the interaction of an $N$-vector real field with a
${\mathbb Z}_2$-gauge field.  It is relevant for transitions in
nematic liquid crystals, see, e.g., Refs.~\cite{LRT-93,KNNSWS-15}, and
in systems with fractionalized quantum numbers, see, e.g.,
Refs.~\cite{SSS-02,SM-02}.  Its cubic-lattice Hamiltonian reads
\begin{eqnarray}
H(J,K) = - J N \sum_{{\bm x},\mu} \sigma_{{\bm x},\mu} \, {\bm s}_{\bm
  x} \cdot {\bm s}_{{\bm x}+\hat{\mu}} + H_G(K), \label{ham}
\end{eqnarray}
where the site variables ${\bm s}_{\bm x}$ are unit-length
$N$-component real vectors, $\sigma_{{\bm x},\mu}\pm 1$ are the link
variables, and $H_G$ is the Hamiltonian of the 
${\mathbb Z}_2$-gauge model with Hamiltonian~(\ref{Hgz2}). 
By measuring energies in
units of the temperature $T$, we can formally set $T=1$ and write the
partition function as
\begin{equation}
Z(J,K)=\sum_{\{{\bm s},\sigma\}} e^{-H(J,K)}.
\label{partfunc}
\end{equation}
For $N=1$ the spin variables take the integer values $s_{\bm x}=\pm
1$, and the model corresponds to the so-called ${\mathbb Z}_2$-gauge
Higgs model~\cite{Wegner-71, BDI-74, FS-79,Kogut-79}.

The Hamiltonian (\ref{ham}) is invariant under global SO($N$)
transformations ${\bm s}_{\bm x} \to V {\bm s}_{\bm x}$ with $V\in
\mathrm{SO}(N)$, and local ${\mathbb Z}_2$-gauge transformations,
i.e., ${\bm s}_{\bm x}\to w_{\bm x} {\bm s}_{\bm x}$ and $\sigma_{{\bm
    x},\nu}\to w_{\bm x} \sigma_{{\bm x},\nu} w_{{\bm x}+\hat{\nu}}$
with $w_{\bm x}=\pm 1$.  Due to the ${\mathbb Z}_2$-gauge invariance,
the correlation function $\langle {\bm s}_{\bm x} \cdot {\bm s}_{\bm
  y} \rangle$ trivially vanishes for ${\bm x}\neq {\bm y}$ and any
value of the Hamiltonian parameters $K$ and $J$.  Therefore, these
correlations cannot characterize the disorder-order DD-O and DO-O
transitions, see Fig.~\ref{phadiaN}, unless a gauge fixing is
applied~\cite{BPV-24-onstar}.  The spontaneous breaking of the global
SO($N$) symmetry is instead signaled by the condensation of the
gauge-invariant bilinear spin-two operator
\begin{eqnarray}
  Q^{ab}_{\bm x} = s_{\bm x}^a s_{\bm x}^b - {1\over N}\delta^{ab}.
  \label{qab}
\end{eqnarray}
The phase diagram of the ${\mathbb Z}_2$-gauge $N$-vector models for
$N\ge 2$ is discussed in Refs.~\cite{FS-79,Sachdev-19,BPV-24-z2gaugeN}
and sketched in Fig.~\ref{phadiaN}.  It shows different phases
characterized by the spontaneous breaking of the global SO($N$)
symmetry and by the different topological properties of the ${\mathbb
  Z}_2$-gauge correlations.  In the following we focus on the
two-component case, i.e., on the ${\mathbb Z}_2$-gauge XY model.

The DD-DO line, that starts on the $J=0$ axis where the model reduces
to the ${\mathbb Z}_2$-gauge model (\ref{Hgz2}), is given
by~\cite{BPV-24-z2gaugeN}
\begin{equation}
  K_c(J)= K_c(J=0) - N J^4 +O(J^6),
  \label{kcj}
\end{equation}
with $K_c(J=0) = 0.761413292(11)$, see Eq.~(\ref{Kcest}).  Transitions
are continuous, in the same universality class as that of the
${\mathbb Z}_2$-gauge model \cite{Wegner-71,FS-79,Sachdev-19,BPV-25}.
Therefore, these transitions are topological without a local order
parameter.  Note however that, unlike the pure ${\mathbb Z}_2$-gauge
model, the Wilson loop does not provide an order parameter, because for $J>0$
it satisfies the perimeter law for any $K$, due to the
screening of the matter field.

Unlike models with $N\ge 3$, the model with $N=2$ undergoes continuous
transitions along all three transition lines of the phase
diagram~\cite{BPV-24-z2gaugeN}. They belong to three of the four
classes outlined in Ref.~\cite{BPV-25} and also reported in
Sec.~\ref{intro}.  The DD-O continuous transitions are LGW
transitions. The order-parameter field is obtained by coarse graining
the gauge-invariant operator (\ref{qab}). DO-O transitions are instead
LGW$^\times$ ones, the order-parameter field being the coarse-grained
gauge-dependent spin variable ${\bm s}_{\bm x}$.  Note that, although
DD-O and DO-O continuous transitions both belong to the XY
universality class, the relevant critical modes are not the same.
Indeed, for small values of $K$ the XY order parameter is associated
with the gauge-invariant operator $Q^{ab}_{\bm x}$ defined in
Eq.~(\ref{qab}), while for large values of $K$, the model has a
nongauge-invariant order parameter, the spin ${\bm s}_{\bm x}$, that
emerges only when an appropriate gauge fixing is
introduced~\cite{BPV-24-z2gaugeN,BPV-24-onstar}.  Along the DD-O and
DO-O lines the correlation-length exponent $\nu$, as well as the
correction-to-scaling exponent $\omega$, are the same as in the XY
model. In the following we use the accurate estimates $\nu=0.6717(1)$
and $\omega=0.789(4)$, see, e.g.,
Refs.~\cite{CHPV-06,Hasenbusch-19,CLLPSSV-20}.

As a consequence of the different nature of the critical modes along
the DD-O and DO-O lines, the critical behavior of the correlations of
the operator $Q_{\bm x}^{ab}$ differs in the two cases. Along the
small-$K$ DD-O transition line, the RG dimension $Y_Q$ of $Q_{\bm
  x}^{ab}$ coincides with the RG dimension $Y_{V,{\rm
    XY}}=0.519088(22)$ of the vector field in the XY universality
class.  On the other hand, along the large-$K$ DO-O line, since the
order parameter is the spin $s_{\bm x}$, $Q_{\bm x}^{ab}$ behaves as a
tensor spin-2 operator; therefore, $Y_Q = Y_{T,{\rm XY}}$, where
$Y_{T,{\rm XY}}=1.23629(11)$ is the spin-two RG dimension in the XY
universality class.  The different values of $Y_Q$ along the two
transition lines have been confirmed numerically~\cite{BPV-24-z2gaugeN}.

\section{Definitions of the dynamic observables}
\label{anreldyn}

To determine the critical exponent $z$, we consider time scales $\tau$ defined
in terms of autocorrelation functions of lattice observables, using the
definitions reported in Refs.~\cite{HPV-07,Hasenbusch-20}.

Given a (gauge-invariant) observable $O$, we consider 
its connected autocorrelation function
\begin{equation}
C_O(t) = \langle O(t_0) O(t_0+t)\rangle_c,
\end{equation}
and several time scales, which are expected to scale as $\xi^z$ in the
critical regime. We first define the self-consisistently truncated
integrated autocorrelation time.  If
\begin{equation}\label{eq:tauint_trunc}
I(t_{\rm max})=\frac{1}{2}+\frac{1}{C_O(0)}\sum_{t=1}^{t_{\rm
    max}}C_O(t)\ ,
\end{equation}
we can define $\tau_{x,{\rm int}}$ self-consistently as the solution 
of the relation
\begin{equation}
\tau_{x,{\rm int}}=I(x\tau_{x,{\rm int}})\ ,
\label{tauint_x}
\end{equation} 
where $x>1$ is a fixed number and 
$I$ has been linearly extended to the whole real line.
The integrated autocorrelation time that 
should be used in the error analysis is obtained in the limit 
$x\to \infty$. For this purpose one should add 
the neglected tail, $\sum_{t=t_{{\rm max}}+1}^{\infty} C_O(t)/C_O(0)$
with $t_{\rm max} = x\tau_{x,{\rm int}}$, 
to the estimate 
$\tau_{x,{\rm int}}$ for a large value of $x$,
as done in Ref.~\cite{Hasenbusch-20}.

One can also define an effective time scale $\tau$ by considering 
a self-consistent finite-time 
exponential autocorrelation time. We define an effective time scale
\begin{equation}\label{taueff}
\tau_{\rm eff}(t+n/2)=\frac{n}{\log\big[C_O(t)/C_O(t+n)\big]}.
\end{equation}
The timescale $\tau_x$ is the solution
of the implicit equation
\begin{equation}
\tau_x=\tau_{\rm eff}(x\,\tau_x)\ ,
\end{equation}
where $x$ is a positive number and we use a linear interpolation to extend the
function $\tau_{\rm eff}$ to the whole real line.  The value of $n$ can be kept
fixed as a function of $L$ or it can be determined self-consistently, keeping
$n/\tau_x$ approximately constant. This second choice is typically more
convenient as it provides estimates with smaller errors, see the discussion in
Ref.~\cite{HPV-07}.

It turns out that the exponential time scale $\tau_x$ is less affected by
scaling corrections than $\tau_{x,{\rm int}}$, especially when autocorrelation
times are not very large; see App.~\ref{data} for some numerical data. In this
work we mostly analyze $\tau_x$ to determine the dynamic critical exponent $z$.
Statistical errors of the autocorrelation time $\tau_x$ have been estimated by using
the blocked jackknife method, with each block consisting of data coming from 
one of the O(50) independent simulations 
performed.

As for the  observable $O$, in the 
present study we consider the energy density, the Polyakov loop
\begin{equation}
P=\frac{1}{L^2}\sum_{x_1, x_2}\prod_{x_3}\sigma_{{\bm x}, 3}\ ,
\end{equation}
where ${\bm x}=(x_1,x_2,x_3)$, and (for the ${\mathbb Z}_2$-gauge
XY model)
the susceptibility of the operator $Q^{ab}$ defined in Eq.~\eqref{qab}
\begin{equation}
\chi_Q=\frac{1}{L^3}\sum_{{\bm x},{\bm y},a,b}Q_{\bm x}^{ab}Q_{\bm y}^{ba}\ .
\end{equation}

\section{Critical relaxational dynamics of the ${\mathbb Z}_2$-gauge model}
\label{reldynz2gau}

In this section we report our numerical analysis of the critical dynamics in
the 3D ${\mathbb Z}_2$-gauge model in equilibrium condition.  Critical dynamic
phenomena crucially depend on the type of dynamics that drives the evolution of
the system. We consider a purely relaxational dynamics without any conservation
law (also known as model A~\cite{HH-77,Ma-book,FM-06}), which can be realized
by using a Langevin or a Metropolis dynamics ~\cite{Binder-76}. More
specifically, we consider a standard (locally reversible) Metropolis dynamics,
in which sites ${\bm x}$ are visited sequentially, in lexicographic order, and
for each site ${\bm x}$ we propose an update of all variables $\sigma_{{\bm
x},\mu}$, in order of increasing $\mu$ value. The time unit corresponds to a
complete sweep of the lattice, i.e., to a single proposed update of all lattice
variables.


At the critical point 
the  time scale $\tau$ of the critical modes is expected to
diverge as~\cite{Ma-book,HH-77,FM-06,HPV-07}
\begin{equation}
  \tau = c \, L^z \left[ 1 + c_\omega\,L^{-\omega} + ...\right],
  \label{taubeh}
\end{equation}
where $L$ is the lattice size, $z$ is the universal exponent
associated with the relaxational dynamics, $c$ and $c_\omega$ are
nonuniversal constants, that also depend on the particular definition
of autocorrelation time, $\omega$ is the universal leading
scaling-correction exponent, and the dots indicate additional
subleading scaling corrections.

We performed simulations of the 3D $\mathbb{Z}_2$ gauge model (\ref{Hgz2}) with
periodic boundary conditions, at the critical point $K_c$ [we use the
estimate~(\ref{Kcest})]. We consider lattices of linear size $L$, with
$8\le L \le 44$, collecting a statistics of the order of $10^9$
lattice sweeps in each case. We computed the effective exponential
autocorrelation time $\tau_x$, defined in Sec.~\ref{anreldyn}, of the
energy and the Polyakov loop, for several values of the parameters $x$
and $n$, entering the definition of $\tau_{\rm eff}$, see
Eq.~\eqref{taueff}.  The autocorrelation times for the Polyakov loop
are significantly larger than those for the energy density (see
App.~\ref{data}), and thus the Polyakov loop appears to be the
observable that is better coupled with the slowest modes of the
dynamics. This is confirmed by the results of the analyses presented
below.  The Polyakov-loop results are affected by significantly
smaller scaling corrections than the results obtained analyzing the
energy density.  For this reason, in the following we will present
only Polyakov-loop data. Results obtained by using energy density
provide consistent results, but with larger uncertainties.

\begin{figure}[tbp]
\includegraphics[width=0.9\columnwidth, clip]{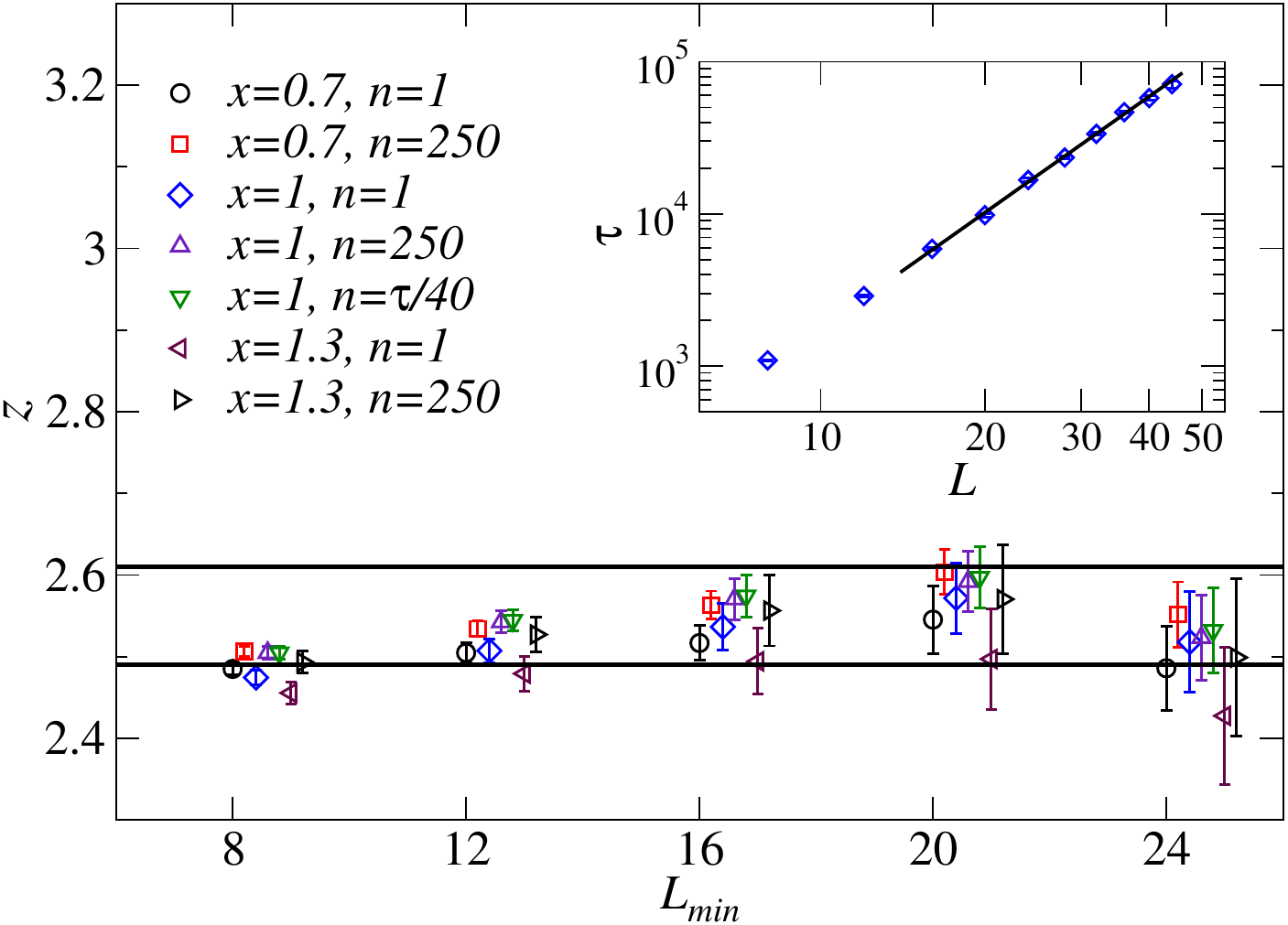}
\caption{Estimates of $z$ obtained from fits of $\tau_{x}$ for several
  values of the parameters $x$ and $n$ (see Sec.~\ref{anreldyn} for
  their definitions), using the Polyakov loop and including only data
  obtained for lattices of linear size $L\ge L_{\rm min}$.  Scaling
  corrections have not been included.  The horizontal band corresponds
  to the final estimate $z=2.55(6)$. As an example, in the inset we
  show the estimates of $\tau_{x}$ for $x=1$ and $n=1$; the line
  corresponds to $a L^z$, with $z=2.55$.}
\label{tauZ2_nocorr}
\end{figure}

To determine the exponent $z$, we fitted the values of $\tau_x$ to
Eq.~\eqref{taubeh}, without scaling corrections, i.e., setting $c_{\omega}=0$.
To estimate the systematic error, the fit was repeated discarding data for
lattices of size $L < L_{\rm min}$, for values of $L_{\rm min}$ varying from
$8$ up to $24$. The results obtained for $z$, 
using different values of $n$ and $x$ to estimate $\tau_{\mathrm{eff}}$, are reported in
Fig.~\ref{tauZ2_nocorr}.  The values of $\tau_x$ are quite insensitive to
the specific choice of $n$ and $x$, and also error bars have quite a mild
dependence on these numbers. As a consequence, the values of $z$ estimated by
using different values of $n$ and $x$ differ at most by two standard
deviations, and this variability is used as a measure of the systematic
uncertainties of the method.

Results are apparently stable for $L_{\rm
  min}\ge 16$, providing a reliable estimate of $z$.  
As final estimate we report
\begin{equation}
  z = 2.55(6),
  \label{zest}
\end{equation}
where the error takes prudentially into account the variation of $z$ when
varying $L_{\rm min}$.  Consistent results are also obtained from fits
including the $O(L^{-\omega})$ scaling correction,
see Eq.~(\ref{taubeh}). Results for the energy density are also consistent.

We can compare our result (\ref{zest}) with earlier results in the
literature.  An analysis of the equilibrium relaxational dynamics was
previously performed in Ref.~\cite{BKKLS-90}, obtaining $z=2.5(3)$,
which is in agreement with our result. On the other hand, our result
is smaller than the recent estimate $z=2.70(3)$ reported in
Ref.~\cite{XCMCS-18}, obtained by analyzing the out-of-equilibrium
relaxational dynamics when slowly crossing the transition point. We
believe that such discrepancy should be further investigated, because
the dynamic exponents $z$ is expected to be the same
for the equilibrium and the out-of-equilibrium critical dynamics.

\section{Critical dynamics in ${\mathbb Z}_2$-gauge $N$-vector models}
\label{reldynz2nvec}

We now report numerical analyses of the critical dynamics at the
continuous transitions of the 3D ${\mathbb Z}_2$-gauge XY model. As
for the ${\mathbb Z}_2$-gauge model, we consider a Metropolis dynamics
(an example of a purely relaxational dynamics without any conservation
law~\cite{Binder-76}) at the critical point in equilibrium
conditions. 

An update of all the lattice variables (one iteration) is obtained by 
first updating all site variables ${\bm s}_{\bm x}$, 
then all link variables $\sigma_{{\bm x},\mu}$,
using the same scheme
already adopted for the $\mathbb{Z}_2$ gauge theory.
In both sweeps the lattice sites are considered sequentially,
in lexicographic order. 
In the Metropolis update of the spin variables the proposed new spin variable
is chosen uniformly on the unit circle.  As for the ${\mathbb Z}_2$ variables,
we propose $\sigma_{{\bm x},\mu}' = -\sigma_{{\bm x},\mu}$.

We compute the autocorrelation functions of the energy density, of the
bilinear operator (\ref{qab}), and of the Polyakov loop.  The
corresponding autocorrelation times are determined using the same
methods employed for the pure ${\mathbb Z}_2$-gauge model, see
Sec.~\ref{anreldyn} for details.

For later comparisons, it is useful to recall that the purely
relaxational dynamics in 3D $N$-vector models has been already studied
numerically and using field-theoretical methods.  The exponent $z$ has
been computed to three-loop order using the $\varepsilon$-expansion
approach.  If we parametrize the result as~\cite{Ma-book}
\begin{equation}
  z = 2 + c\,\eta,
  \label{zetaon}
  \end{equation}
where $\eta$ is the critical susceptibility exponent, we obtain for
$c$~\cite{FM-06,AV-84,HH-77,HHM-72}
\begin{eqnarray}
  c\approx c_0+ c_1 \varepsilon,\quad
  c_0=6\ln(4/3)-1,\;\;c_1\approx - 0.1369.
  \label{ceps}
  \end{eqnarray}
Note that $c$ is independent of $N$ at order $\varepsilon$. We also
mention that $c=1/2+O(1/N)$ in the large-$N$ limit~\cite{Ma-book}.
Inserting the best estimate of $\eta$ for the XY universality class
into Eq.~(\ref{zetaon}), i.e., $\eta=0.03810(8)$~\cite{Hasenbusch-19},
we obtain $z=2.022(5)$, where, as an estimate of the uncertainty, we
take the contribution of the $O(\varepsilon)$ term of $c$. An
analogous computation for the 3D Ising universality would give
$z=2.021(5)$, in agreement with the more accurate estimate
$z=2.0245(15)$ obtained by numerical analyses of equilibrium MC
simulations~\cite{Hasenbusch-20}.

\subsection{Relaxational dynamics at the topological DD-DO transitions}
\label{relaxtopo}

\begin{figure}[tbp]
\includegraphics[width=0.9\columnwidth, clip]{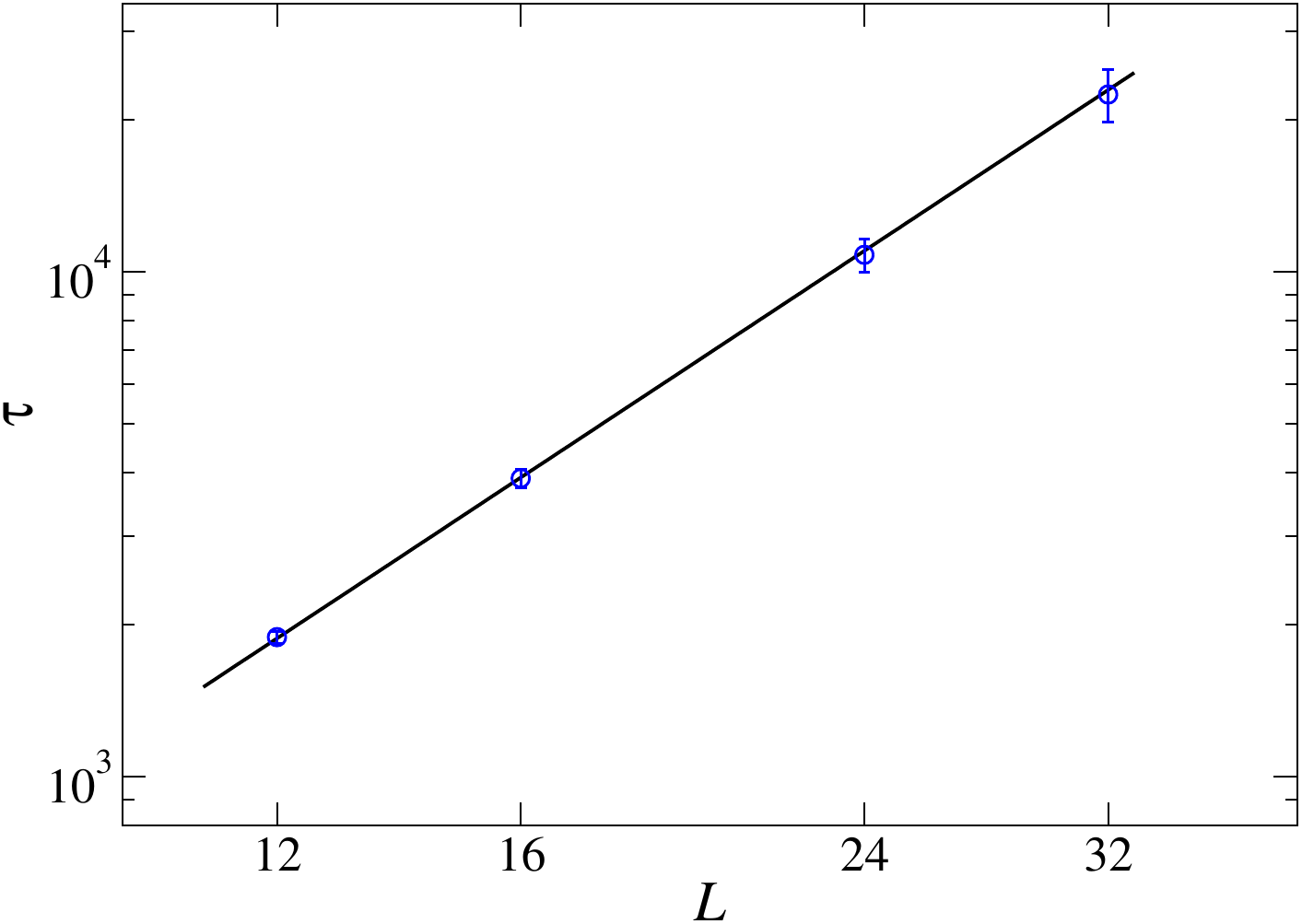}
\caption{Estimates of $\tau_{x,\mathrm{int}}$ with $x=2$, see
  Sec.~\ref{anreldyn}, for the Polyakov loop. The solid line is a fit
  of the data to $\tau=AL^{z}$, with $z$ fixed to $z=2.55$, which is
  our estimate of $z$ for the ${\mathbb Z}_2$-gauge model.}
\label{tauXYJ01}
\end{figure}

To begin with, we consider the DD-DO transitions, which, as already
mentioned, belong to the 3D ${\mathbb Z}_2$-gauge universality
class~\cite{Wegner-71,FS-79,Sachdev-19,BPV-25}.  To investigate the
nature of the relaxational dynamics, we consider the transition along
the line $J=0.1$, which occurs at $K_c\approx 0.7612$ according to
Eq.~(\ref{kcj}).  We perform MC simulations on lattices of linear size
up to $L=32$, determining the autocorrelation functions of the energy
density and Polyakov loop. The analysis of the corresponding
autocorrelation times (see App.~\ref{data} for some results) 
show that the dynamic critical behavior is the
same as that of the ${\mathbb Z}_2$-gauge model discussed in
Sec.~\ref{reldynz2gau}. Indeed, as shown in Fig.~\ref{tauXYJ01}, the
critical behavior of the autocorrelation times is consistent with the
exponent $z=2.55(6)$ determined in the 3D ${\mathbb Z}_2$-gauge model.
An unbiased analysis of the data leads to the consistent result
$z=2.52(8)$.

\subsection{Critical dynamics at the DD-O transitions}
\label{relaxLGW}

Along the DD-O line the ${\mathbb Z}_2$-gauge XY model undergoes XY
transitions~\cite{BPV-24-z2gaugeN}, the order parameter being the
gauge-invariant bilinear operator (\ref{qab}). Therefore, they are LGW
transitions, according to the classification of Ref.~\cite{BPV-25}
reported in the introduction. We note that the Metropolis update
acts on the gauge-dependent variables, therefore its action on the
bilinear field (\ref{qab}) is not direct.

\begin{figure}[tbp]
\includegraphics[width=0.9\columnwidth, clip]{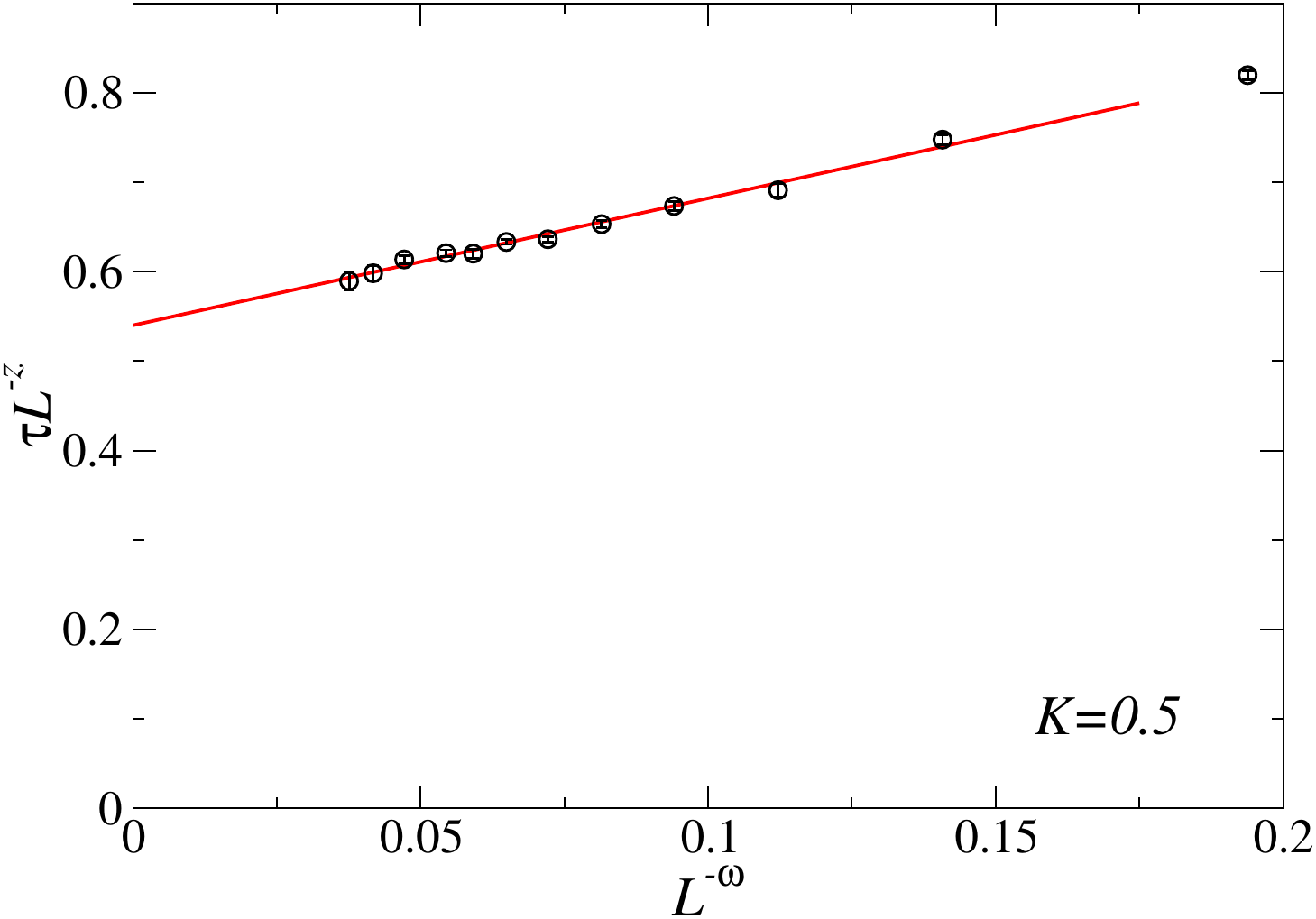}
\caption{Rescaled estimates of the autocorrelation time $\tau_x$ for
  the ${\mathbb Z}_2$-gauge XY model along the DD-O transition line,
  at $K=0.5$ and $J_c=0.37118$, using the XY exponent $z = 2.022$.
  The autocorrelation time has been computed using the susceptibility
  of the bilinear operator $Q^{ab}_{\bm x}$ defined in Eq.~(\ref{qab})
  and fixing $x = 1$, $n\approx \tau_x/40$; see Eq.~\eqref{taueff} for
  the definition of $\tau_x$ 
  and Tab.~\ref{TabDDO} for the data. The solid line corresponds to a linear fit of the
  data for $L\ge 16$ to $\tau = a L^z (1 + c L^{-\omega})$ fixing $z$
  and $\omega$ to the XY values $z=2.022$ and $\omega=0.789$
  ($\chi^2/{\rm d.o.f}\approx 1.3$). }
\label{tauXY05}
\end{figure}

We focus on the DD-O transition along the line $K=0.5$, see
Fig.~\ref{phadiaN}, whose equilibrium critical behavior was already
studied in Ref.~\cite{BPV-24-z2gaugeN}.  The transition point is
located at $J_c=0.37118(2)$. We perform MC simulations at $J_c$ for
several values of $L$, up to $L=64$, collecting a statistics of
roughly $3\times 10^8$ lattice iterations
for the larger lattices. The
most precise estimates of the critical time scale $\tau$ are obtained
from the autocorrelation functions of the susceptibility of the
bilinear operator $Q^{ab}_{\bm x}$ defined in Eq.~(\ref{qab}); 
see App.~\ref{data} for some data.  The
behavior of $\tau$ as a function of $L$ is fully consistent with that
expected for a model in the standard XY universality class. Indeed,
$\tau$ behaves as $\tau\sim L^z$, where $z=2.022(5)$ is the value of
$z$ appropriate for the XY universality class. This can be clearly
seen from Fig.~\ref{tauXY05}, in which we also report the result of a
fit to Eq.~\eqref{taubeh}, with $z$ and $\omega$ fixed to
the values determined for the XY universality class.  If we instead fix
$\omega=0.789$ (the XY value) and keep $z$ as a free parameter, 
we obtain the consistent estimate $z=2.02(4)$.
 
\subsection{Critical dynamics at DO-O transitions}
\label{relaxLGWplus}

Along the DO-O line the ${\mathbb Z}_2$-gauge XY model undergoes XY
transitions. The order parameter is the gauge-dependent spin variable
${\bm s}_{\bm x}$, while the gauge-invariant bilinear operator
(\ref{qab}) behaves as a spin-2 composite
operator~\cite{BPV-24-z2gaugeN}. These transitions belong to the
LGW$^\times$ class.

\begin{figure}[tbp]
\includegraphics[width=0.9\columnwidth, clip]{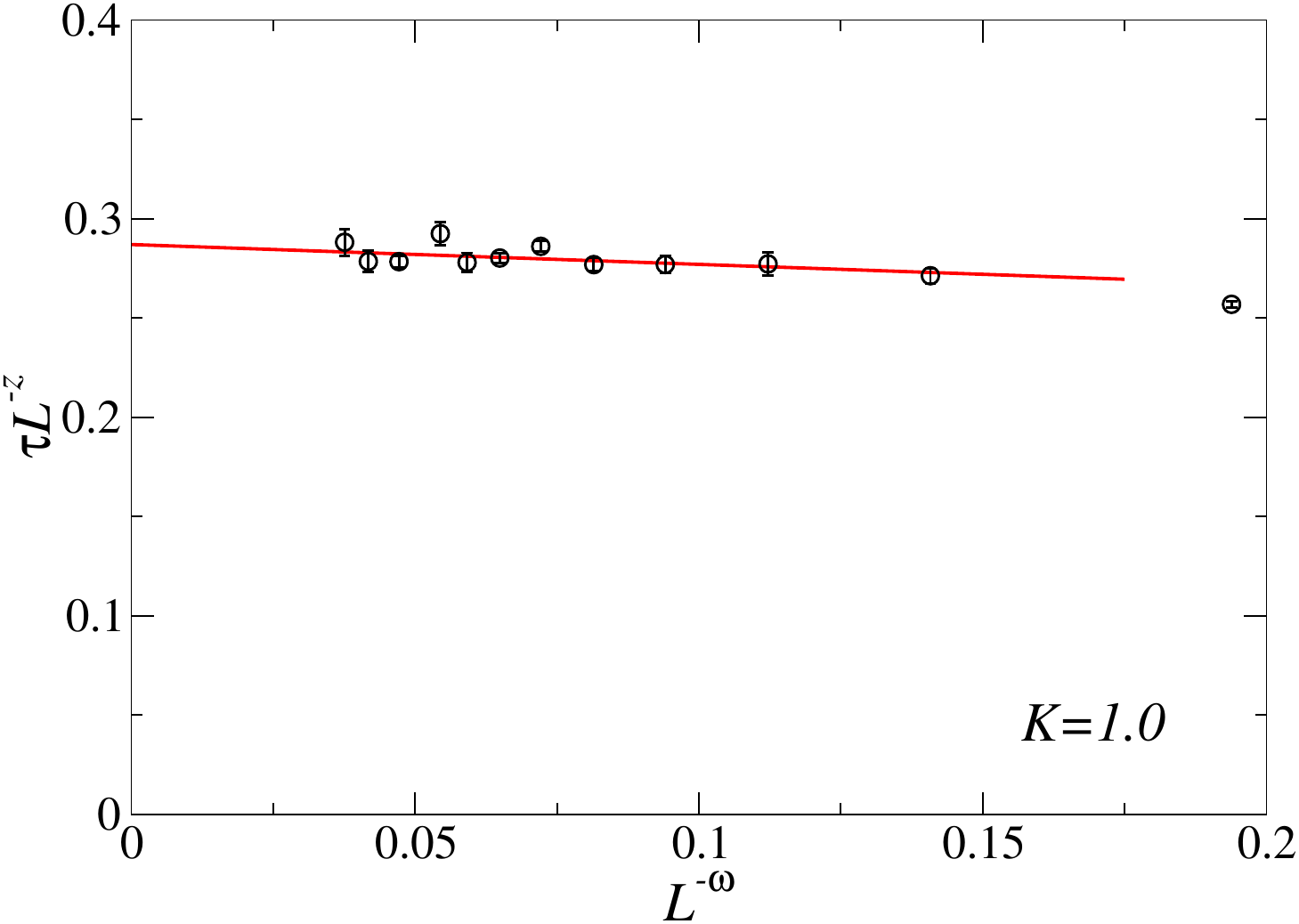}
\caption{Rescaled estimates of the autocorrelation time $\tau_x$ for
  the ${\mathbb Z}_2$-gauge XY model along the DO-O transition line,
  at $K=1$ and $J_c=0.22729$, using the XY exponent $z = 2.022$.  The
  autocorrelation time has been computed using the susceptibility of
  the bilinear operator $Q^{ab}_{\bm x}$ defined in Eq.~(\ref{qab})
  and fixing $x = 1$, $n\approx \tau_x/20$; 
  see Eq.~\eqref{taueff} for
  the definition and Tab.~\ref{TabDOO} for the data. The solid line corresponds to a linear fit of the
  data for $L\ge 16$ to $\tau = a L^z (1 + c L^{-\omega})$ fixing $z$
  and $\omega$ to the XY values $z=2.022$ and $\omega=0.789$
  ($\chi^2/{\rm d.o.f}\approx 1.3$). }
\label{tauXY1}
\end{figure}

To investigate the nature of the relaxational dynamics along the DO-O
line, we focused on the DO-O transition along the line $K=1$, see
Fig.~\ref{phadiaN}.  The static critical behavior along the line was
already studied in Ref.~\cite{BPV-24-z2gaugeN}, finding a continuous
transition at $J_c=0.22729(3)$. As in the previous case, we performed
MC simulations at $J_c$ for several values of $L$, up to $L=64$,
collecting a statistics of roughly $3\times 10^8$ iterations
for the
largest lattices.  The most precise estimates of the critical time
scale $\tau$ were again obtained from the autocorrelation function of
the susceptibility of the gauge-invariant operator defined in
Eq.~(\ref{qab}); see App.~\ref{data} for some estimates. 
The behavior of $\tau$ as a function of the lattice
size $L$ is shown in Fig.~\ref{tauXY1}, and is again in very good
agreement with the power-law behavior $\tau\sim L^z$, where
$z=2.022(5)$ is the standard XY dynamic exponent.  If we instead 
fit the data to $\tau = a L^z (1 + c L^{-\omega})$
fixing $\omega$ to the XY value, $\omega=0.789$, 
we obtain the consistent estimate $z=1.96(8)$.

\section{Conclusions}
\label{conclu}

The critical dynamics of statistical systems with gauge symmetries is
fundamental to understand the mechanisms underlying critical phenomena emerging
in gauge systems.  The critical behavior in the presence of gauge symmetries is
more complex than in standard continuous transitions characterized by a global
symmetry, which can be described by effective LGW $\Phi^4$ theories. In
particular, there are also continuous transitions that cannot be described by
the LGW paradigm. In general, continuous transitions of statistical models in
the presence of gauge symmetries can be classified in four general
classes~\cite{BPV-25}, depending on the role of the gauge and matter fields, of
the global symmetry (if present), and on the nature of the order parameter, as
also discussed in the introduction. It is interesting to extend this
characterization of the static critical behavior to dynamic critical phenomena.

As a first step, we study here the universal features of a Metropolis
local dynamics---this is an example of a purely relaxational dynamics
or model-A dynamics in the classification of Ref.~\cite{HH-77}---at
the phase transitions occurring in the 3D ${\mathbb Z}_2$-gauge model
with Hamiltonian (\ref{Hgz2}) and in the 3D ${\mathbb Z}_2$-gauge XY
model.  As discussed in Sec.~\ref{model}, these lattice gauge models
present different types of continuous transitions.  The transition in
the 3D ${\mathbb Z}_2$-gauge model and the continuous transitions
along the DD-DO line of the 3D ${\mathbb Z}_2$-gauge XY model, see
Fig.~\ref{phadiaN}, are topological. On the other hand, the DD-O and
DO-O transitions in the ${\mathbb Z}_2$-gauge XY model are examples of
LGW and LGW$^\times$ transitions,
respectively~\cite{BPV-25,BPV-24-z2gaugeN}.

The topological continuous transitions in the ${\mathbb Z}_2$-gauge
model and in the ${\mathbb Z}_2$-gauge XY model (DD-DO line) belong to
the ${\mathbb Z}_2$-gauge universality class.  The existence of an
exact duality mapping between the partition functions of the ${\mathbb
  Z}_2$-gauge model and of the standard cubic-lattice Ising
model~\cite{Wegner-71,FS-79,Savit-80,Kogut-79} implies that
energy-related observables have the same static critical behavior in
the two models. On the other hand, our results show that the dynamic
critical behavior differs. The relaxational dynamics in the gauge
model is significantly slower than in the Ising model. Indeed, our
estimate of the dynamic exponent $z$ for the gauge model, $z=2.55(6)$,
is significantly larger than that 
for the Ising universality class,
$z=2.0245(15)$~\cite{Hasenbusch-20}, confirming earlier numerical
results obtained by analyzing out-of-equilibrium critical
behaviors~\cite{XCMCS-18}.  This difference can be explained by noting
that the duality mapping is nonlocal, so the Ising dynamics would be
equivalent to a nonlocal dynamics in the gauge model. The local
dynamics we consider in the 3D ${\mathbb Z}_2$-gauge model is
therefore unrelated and qualitatively different, giving rise to a
different critical dynamic behavior with a different exponent $z$.

Concerning the continuous transitions in the 3D ${\mathbb Z}_2$-gauge
XY model along the DD-O and DO-O lines, see Fig.~\ref{phadiaN}, our
numerical analyses show that the relaxational dynamics belongs to the
same dynamic universality class as the model-A dynamics in the
standard XY universality class~\cite{HH-77}. Indeed, our estimates of
the exponent $z$ are consistent with the estimate $z=2.022(5)$
obtained for the standard XY universality class~\cite{HH-77}.  We
remark that this result is not obvious. Indeed, it implies that gauge
modes do not play any role in the dynamics and also that the nature of
the order parameter is irrelevant.
Moreover, we note that in the case of the LGW transitions along the
DD-O line only gauge-invariant modes are present in the corresponding
LGW $\Phi^4$ theory---the order parameter is the gauge-invariant 
operator $Q^{ab}_{\bm x}$ and not the gauge-dependent spin variable 
${\bm s}_{\bm x}$---and thus only these modes evolve in the
corresponding model-A Langevin dynamics, while in the lattice
gauge model the basic updating rules are defined in terms of the
non-gauge-invariant fundamental variables, $\sigma_{{\bm x},\mu}$ and
${\bm s}_{\bm x}$ in the models we consider here. 

The results obtained for the critical relaxational dynamics at the LGW
and LGW$^\times$ continuous transitions of the 3D ${\mathbb
  Z}_2$-gauge XY model suggest a general scenario for the
critical dynamics at continuous LGW and LGW$^\times$ transitions in
the presence of gauge symmetries. We conjecture that the relaxational
critical dynamics at LGW and LGW$^\times$ transitions belongs to the
same dynamic universality class as the purely relaxational
Langevin dynamics defined in the corresponding LGW $\Phi^4$
theory~\cite{Ma-book,HH-77}.

It is interesting to extend the present analysis to models with
continuous gauge symmetry groups, such as the Abelian U(1) and the
non-Abelian SU($N$) group.  Indeed, these models also present
transitions where the gauge modes become critical---GFT transitions in
the classification of Ref.~\cite{BPV-25}, see also Sec.~\ref{intro}.
Investigations of their critical relaxational dynamics may provide
further insights in the mechanisms working at the transition lines
bounding Higgs phases.

\acknowledgments

The authors acknowledge support from project PRIN 2022 ``Emerging
gauge theories: critical properties and quantum dynamics''
(20227JZKWP). Numerical simulations have been performed on the CSN4
cluster of the Scientific Computing Center at INFN-PISA.

\appendix

\section{Some numerical data}
\label{data}

In this appendix we report some values of the autocorrelation times
estimated along the different transition lines.

\begin{table}[tbh]
\begin{tabular}{lll|ll} 
\hline \hline  
$L$ & $n$ & $\tau_x(H)$ & $n$ & $\tau_x(P)/10^4$ \\ \hline 
8  & 1 & 20.91(5) & 25 & 0.1090(8) \\  
12 & 1 & 57.2(4) & 70 & 0.290(3)  \\
16 & 3 & 119.9(1.3) & 145 & 0.593(13)  \\
20 & 5 & 212(2) & 245 & 1.00(3) \\ 
24 & 8 & 326(8) & 410 & 1.77(5) \\ 
28 & 12 & 491(7) & 580 & 2.41(5) \\
32 & 16 & 714(16) & 840 & 3.49(6)\\
36 & 23 & 947(16) & 1150 & 4.79(8) \\   
40 & 28 & 1231(25) & 1400 & 6.15(14) \\ 
44 & 34 & 1680(100) & 1700 & 7.85(3) \\ \hline\hline
\end{tabular}
\caption{Autocorrelation times of the 3D ${\mathbb Z}_2$-gauge model at the
  critical point $K_c$ [we use the numerical value is reported in 
  Eq.~(\ref{Kcest})]. We report the values of
  $\tau_x$, defined in Eq.~\eqref{taueff},
   for the energy density ($H$) and the Polyakov loop ($P$), 
  setting $x=1$ and $n\approx \tau_x/40$.
  Data are in units of lattice iterations, i.e.,  
  one Metropolis proposed update of all lattice link variables.}
\end{table}

\begin{table}[tbh]
\begin{tabular}{lll} 
\hline \hline $L$ & $\tau_x(P)/10^3$ & $\tau_{x, int}(P)/10^3$ \\ \hline 
12   & 3.0(3)   &  1.89(5)  \\ 
16   & 6.64(36) &  3.90(16) \\ 
24   & 15(2)    &  10.8(8)  \\ 
32   & 39(7)    &  22(3)    \\ \hline\hline
\end{tabular}
\caption{Autocorrelation times for the Polyakov loop in the 3D
  ${\mathbb Z}_2$-gauge XY model at $J=0.1$, $K=0.7612$.  We report
  the values of $\tau_x$, defined in Eq.~\eqref{taueff}, for $x=1$ and
  $n=512$, and of $\tau_{x,\rm int}$, defined in Eq.~\eqref{tauint_x},
  for $x=2$.  Data are in units of lattice iterations, i.e., one
  Metropolis proposed update for all lattice variables.}
\end{table}

\begin{table}[tbh]
\begin{tabular}{lll} 
\hline \hline
$L$ & $n$ & $\tau_x(\chi_Q)$      \\ \hline 
8   & 1   &  54.9(3)    \\
12  & 2   &  113.7(8)   \\
16  & 4   &  188(2)   \\ 
20  & 6   &  288(2)   \\ 
24  & 9   &  403(3)   \\ 
28  & 12  &  537(3)   \\ 
32  & 16  &  700(3)   \\ 
36  & 20  &  870(7)   \\ 
40  & 25  &  1077(6)  \\ 
48  & 36  &  1539(11)     \\ 
56  & 49  &  2049(28)     \\ 
64  & 64  &  2645(44)     \\ \hline\hline
\end{tabular}
\caption{Autocorrelation times for the 3D ${\mathbb Z}_2$-gauge XY
  model at $K=0.5$, $J=0.37118$. Values of $\tau_x$ for $x=1$ and
  $n\approx \tau_x/40$, see Eq.~\eqref{taueff}, for the susceptibility
  of the order parameter $Q^{ab}$ defined in Eq.~\eqref{qab}.  Data
  are in units of lattice iterations; a lattice iteration consists in a
  proposed update of all $\sigma_{{\bm x},\mu}$ and ${\bm s}_{\bm x}$
  variables.  Autocorrelation times obtained by using the energy
  density are quite similar to those obtained from the susceptibility
  of $Q^{ab}$.}
\label{TabDDO}
\end{table}

\begin{table}[tbh]
\begin{tabular}{lll} 
\hline\hline
$L$ & $n$ & $\tau_x(\chi_Q)$      \\  \hline
8   & 1   &   17.21(11)  \\ 
12  & 2   &   41.2(6)  \\   
16  & 4   &   75(2)  \\     
20  & 6   &   118(2) \\     
24  & 9   &   171(2) \\     
28  & 12  &   241(2) \\     
32  & 16  &   310(3) \\     
36  & 20  &   390(7) \\     
40  & 25  &   507(10)    \\ 
48  & 36  &   698(7) \\     
56  & 49  &   954(18)    \\ 
64  & 64  &   1293(30)   \\  \hline\hline
\end{tabular}
\caption{Autocorrelation times for the 3D ${\mathbb Z}_2$-gauge XY
  model at $K=1$, $J=0.22729$. Values of $\tau_x$ for $x=1$ and
  $n\approx \tau_x/20$, see Eq.~\eqref{taueff}, for the susceptibility
  of the order parameter $Q^{ab}$ defined in Eq.~\eqref{qab}.  Data
  are in units of lattice iterations; a lattice iteration consists in a
  proposed update of all $\sigma_{{\bm x},\mu}$ and ${\bm s}_{\bm x}$
  variables.  Autocorrelation times obtained by using the energy
  density are quite similar to those obtained from the susceptibility
  of $Q^{ab}$.}
\label{TabDOO}
\end{table}

\end{document}